\documentclass[aps,pra]{revtex4}
\usepackage{graphicx}
\usepackage{hyperref}

\begin{document}

\title{Local density approximation for a perturbative equation of state}
\author{G. E. Astrakharchik}
\affiliation{Dipartimento di Fisica, Universit\`a di Trento and BEC-INFM, I-38050 Povo, Italy\\
Institute of Spectroscopy, 142190 Troitsk, Moscow region, Russia}
\date{\today}

\begin{abstract}
The knowledge of a series expansion of the equation of state provides a deep insight
into the physical nature of a quantum system. Starting from a generic
``perturbative'' equation of state of a homogeneous ultracold gas we make
predictions for the properties of the gas in the presence of harmonic confinement.
The local density approximation is used to obtain the chemical potential, total and
release energies, Thomas-Fermi size and density profile of a trapped system in
three-, two-, and one- dimensional geometries. The frequencies of the lowest
breathing modes are calculated using scaling and sum-rule approaches and could be
used in an experiment as a high precision tool for obtaining the expansion terms of
the equation of state. The derived formalism is applied to dilute Bose and Fermi
gases in different dimensions and to integrable one-dimensional models. Physical
meaning of expansion terms in a number of systems is discussed.
\end{abstract}

\maketitle

\section{Introduction}

Recent progress in experimental techniques has made it possible to investigate
the properties of gases of bosonic atoms in traps at temperatures lower than that
required to achieve
Bose-Einstein condensation (BEC), as well as the properties of
fermionic atoms at temperatures lower than the Fermi temperature. When the density
of a homogeneous gas is small enough perturbation theory can be applied and the
equation of state of repulsive bosons and two-component fermions can be expressed in
terms of the expansion parameter $na^3_s$, where $n$ is the density and $a_s$ is the
$s$-wave scattering length. The problem of a cold trapped gas is more difficult as
another parameter, the harmonic oscillator length $a_{ho}$, enters into the problem.
Although many properties of a homogeneous system (equation of state, {\it etc.}) can
be predicted quite easily, no exact solutions in the presence of an external
potential are generally known. If the number of particles is sufficiently large and
the energy per particle is much larger than the level spacing of the harmonic
oscillator, one can apply the local density approximation (LDA). This method is
commonly and successfully used to relate properties of an inhomogeneous system to
the ones of the corresponding homogeneous system. In this way the predictions of
known homogeneous theory can be tested experimentally in confined systems.

Alternatively, the problem can be seen from the other side. It might happen that it
is possible to observe and investigate properties of a trapped system in an
experiment, while an analytical description of the same physical system is not known
even for a homogeneous case. Then the experimental study of a trapped system might
shed light on the properties of a homogeneous system and allow us to judge the
validity of the analytical approaches. For example, this is the case with the
two-component Fermi gas in the BEC-BCS crossover. The unitarity regime was already
observed in experiments (see, for example, \cite{Bartenstein04b,Kinast05}), while
there is a lack of a well-posed and complete theory. Another intriguing question is
presence or absence of the Lee-Yang term, corresponding to a beyond mean field term
of composite bosons, in the equation of state on the BEC side where formation of
molecules was observed \cite{Jochim03,Greiner03,Zwierlein03}. This problem is still
an open question and there is no conclusive theoretical answer. A high precision
experiment might resolve the question.

The aim of the current work is to investigate how the expansion of the equation of
state of a homogeneous system shows up in properties of a trapped system. In order
to do so we assume that the equation of state of a homogeneous gas can be written as
as series expansion in some limit. The local density approximation is used to obtain
measurable energetic and spatial properties of a trapped system as a function of the
parameters of the expansion and LDA parameter. Special attention is paid to the
prediction of the frequencies of collective oscillations, as a sensitive
experimental tool for investigation of the equation of state. Scaling and sum-rule
approaches are used for deriving the frequencies of small collective oscillations.
The derived formalism is applied to dilute Bose and Fermi gases in different
dimensions and to integrable one-dimensional models. We discuss the physical meaning
of expansion terms in a number of systems.

The paper is organized as follows. In Section~\ref{secLDA} we explain the local
density approximation in an arbitrary number of dimensions and introduce dimensionless
units. In Sec.~\ref{secSolution} we introduce the ``perturbative'' equation of state
and solve the local density approximation problem. Explicit expressions for the
chemical potential, energy, release energy, size and density profile of a trapped gas
are obtained. In Sec.~\ref{secFreq} we use scaling and sum-rule approaches for the
prediction of the frequencies of collective oscillations. Cases of spherical,
elongated, ``cigar''- and ``pancake''- shaped traps are considered. Formulas are
derived for three-, two- and one- dimensional traps. The LDA is applied to relate
the ``perturbative'' equation of state to frequencies of the breathing mode. In
Sec.~\ref{secDiscussion} we apply the developed formalism to a number of physical
systems where perturbation expansion of the equation of state is known. These
include dilute bosonic gases (in 3D, quasi 2D, and quasi 1D geometries),
two-component Fermi gases (in 3D and in 1D), as well as a number of integrable
one-dimensional models (Lieb-Liniger, Calogero-Sutherland). In
Sec.~\ref{secConclusions} we draw our conclusions.

\section{Local density approximation\label{secLDA}}

We consider a gas of $N$ particles (bosons or fermions) at a density $n=N/V$. We
suppose that the density dependence of the chemical potential $\mu_{hom}$ of the
homogeneous system is known from either perturbation theory or from a fit to
experimental data. Within the {\it local density approximation} one assumes (see,
for example, \cite{Pitaevskii03}) that the chemical potential $\mu$ of a trapped
system is given by the sum of the local chemical potential, which is taken to be
equal to the chemical potential of the uniform system $\mu_{hom}$ at the
corresponding density, and the external field $V_{ext}$:
\begin{eqnarray}
\mu = \mu_{hom}(n(\mathbf{r}))+V_{ext}(\mathbf{r})  \label{LDA}
\end{eqnarray}

In the following we will always consider a harmonic external confinement for two
basic reasons. First, this is the most widely used trapping potential in cold
gases. Second, a quadratic potential can be used to describe the external potential
close to its minimum. In order to consider the most general case we will assume that the
trapping frequencies are
different, so that the confinement is anisotropic $V_{ext}(\mathbf{r})=\frac{
1}{2}m\omega_xx^2+\frac{1}{2}m\omega_yy^2+\frac{1}{2}m\omega_zz^2$.

The value of the chemical potential $\mu$ of a trapped gas is fixed by the
normalization condition
\begin{eqnarray}
N=\int \!n(\mathbf{r}){dr}=\int \!\!\!\!\int \!\!\!\!\int \mu_{\hom}^{-1}
\left[ \mu -\frac{1}{2}m\omega_xx^2-\frac{1}{2}m\omega_yy^2-\frac{1}{2}
m\omega_zz^2\right] dxdydz  \label{LDAN}
\end{eqnarray}
Here the density profile is obtained by inverting the density dependence of
the local chemical potential $n=\mu_{\hom}^{-1}$.

The spatial extensions of the cloud, $R_x,R_y,R_z$, depend on the value of
the chemical potential and frequencies of the harmonic confinement:
\begin{eqnarray}
\mu =\frac{1}{2}m\omega_x^2R_x^2=\frac{1}{2}m\omega _y^2R_y^2=\frac{1}{2}
m\omega_z^2R_z^2  \label{LDAR}
\end{eqnarray}

By expressing the distances in the trap in units of the size of the cloud:
$\tilde{r}=(x/R_x,y/R_y,z/R_z)$ it follows immediately that the frequencies of the
trap enter in  Eq.~(\ref{LDAN}) only through the combination
$\omega_{ho}=(\omega_x\omega_y\omega_z)^{1/3}$ and the normalization condition can
be conveniently written in terms of the averaged oscillator length
$a_{ho}=\sqrt{\hbar /m\omega_{ho}}$. Now the integral is to be taken inside a sphere
of radius $1$ and is symmetric with respect to $\tilde{r}$.

The normalization condition (\ref{LDAN}) can be conveniently rewritten in
dimensionless units:
\begin{eqnarray}
\Lambda_{D}^{D/2}=\tilde{\mu}^{D/2}\int \mu_{\hom}^{-1}\left[ \tilde{\mu}
\Lambda_{D}(1-\tilde{r}^2)\right] \mathbf{d\tilde{r}},
\qquad
n(\mathbf{\tilde{r}}) = 0 \text{~for~} |\mathbf{\tilde{r}}|>1
\label{LDANdim}
\end{eqnarray}
Here the chemical potential $\tilde{\mu}$ is expressed in units of
$\frac{1}{2}N^{1/D}\hbar\omega_{ho}$, which is a natural unit of energy in the trap,
the density in a homogeneous system $\mu_{\hom}^{-1}$ is measured in units of
$a^{-D}$, where $a$ is a length scale convenient for the homogeneous system (for
example, it can be equal to the $s$-wave scattering length $a_s$), chemical
potential ({\it i.e.} the argument of the inverse function $\mu_{\hom}^{-1}$) is
measured in units of $\hbar^2/ma^2$, $D=1,2,3$ is the dimensionality of the system,
and the LDA characteristic parameter $\Lambda_D$ is defined as
\begin{eqnarray}
\Lambda_D=N^{1/D}\frac{a^2}{a_{ho}^2}  \label{Lambda}
\end{eqnarray}

From the Eq.~(\ref{LDANdim}) it becomes clear that within the LDA there is a scaling
in terms of the characteristic parameter $\Lambda_{D}$. In other words systems
having a different number of particles and different frequencies of the harmonic
confinement will have the \emph{same} density profile and other LDA properties (once
expressed in appropriate units as discussed above) if they have equal values of the
parameter $\Lambda_D$.

\section{Solution for ``perturbative'' equation of state\label{secSolution}}

In this Section we develop a general theory for a ``perturbative'' equation of state
in a three-, two-, and one- dimensional systems. We start by writing down the
equation of state $\mu_{\hom}(n)$. Assuming that a perturbation approach can be
applied to a homogeneous system, we suppose that $\mu_{\hom}(n)$ is known as a
series expansion, either at small or at high density, in terms of the expansion
parameter (it can be gas parameter $na^D$ in the weakly interacting regime and
$1/(na^D)$ in the regime of strong interactions):
\begin{equation}
\mu_{\hom} = C_1(na^D)^{\gamma_1}(1+C_2(na^D)^{\gamma_2}+...)\frac{\hbar^2}{ma^2},
\label{muhom}
\end{equation}

In this formula the second term is considered to be small compared to the first term
($|C_2(na^3)^{\gamma_2}|\ll 1$) and will be treated perturbatively in the following.
We will limit the series to the second term, as it is responsible for the main
contribution to the dependence of the frequency of the breathing mode on the
density, and we omit higher terms in the expansion~(\ref{muhom}).

We obtain the value of the chemical potential of a trapped system by resolving the
normalization condition~(\ref{LDANdim}).
The result can be conveniently expressed in terms of the characteristic parameter
$\Lambda_D$ (\ref{Lambda}).
\begin{equation}
\tilde{\mu}\equiv\frac{\mu }{\frac{1}{2}N^{1/D}\hbar\omega_{ho}}
=C_{\mu 1}\Lambda_D^{\frac{D\gamma_1-2}{D\gamma_1+2}}\left(1+C_{\mu 2}{
\Lambda}_{D}^{\frac{2D\gamma_2}{D\gamma_1+2}}+...\right),
\label{mu}
\end{equation}
where constants $C_{\mu 1}$, $C_{\mu 2}$ are independent of the parameters of the
trap and, instead, are directly related to the equation of state of a homogeneous
system. In order to find them we evaluate the integral~(\ref{LDANdim}) in three-,
two-, and one- dimensional geometries. The explicit expressions for the coefficients
$C_{\mu 1}$ and $C_{\mu 2}$ are given in Table~\ref{tableCoef}. The leading term in
(\ref{mu}) depends only on two parameters, $C_1$ and $\gamma_1$, of the homogeneous
equation of state, while in order to obtain the next term one has to know four
parameters: $C_1$, $\gamma_1$, $C_2$, $\gamma_2$. This type of behaviour is quite
general and applies to majority of the quantities we estimate in a trapped gas,
with the exception of the frequencies of collective oscillations, where the leading
term depends only on $\gamma_1$ (see below).

\begin{table}[ht!]
\centering
\begin{tabular}{|c|c|c|}
\hline
dimensionality & $C_{\mu 1} $& $C_{\mu 2}$\\ \hline
3D &
$2{\left(\frac{{{C_1^{\frac{1}{\gamma_1}}}}}{{\left(2\pi
\right) }^{\frac{3}{2}}}\frac{\Gamma\left(\frac{5}{2}+\frac{1}{\gamma_1}
\right) }{\Gamma\left(1+\frac{1}{\gamma_1}\right) }\right) }^{\frac{
2\gamma_1}{3\gamma_1+2}}$ &
$\frac{{\left(C_{{\mu 1} }\pi\right)}^{\frac{3}{2}}{\left(\frac{C_{\mu
1}}{2C_1}\right)}^{\frac{1+\gamma_2}{{{\gamma }_{1}}}}\Gamma\left(1+\frac{1+\gamma_2}{
\gamma_1}\right){C_2}}{\left(1+\frac{3}{2}\gamma_1\right)\Gamma
\left(\frac{5}{2}+\frac{1+\gamma_2}{\gamma_1}\right)}$\\\hline
2D&
${\left(2C_1\right)}^{\frac{1}{1+\gamma_1}}{\left(\frac{1+\gamma_1}{\pi\gamma_1}
\right)}^{\frac{\gamma_1}{1+\gamma_1}}$&
$\frac{{C_2}}{1 +\gamma_1 +\gamma_2}{\left(\frac{1 +\gamma_1}{2\pi{C_1}\gamma_1}
\right)}^{\frac{\gamma_2}{1 +\gamma_1}}$\\\hline
1D&
$\frac{{1}}{\pi }{\left(\frac{{\left( 2\pi{C_1}\right)}
^{\frac{1}{\gamma_1}}\Gamma\left(\frac{3}{2}+\frac{1}{{\gamma_1}}\right)}
{\Gamma\left(1+\frac{1}{\gamma_1}\right)}\right)}
^{\frac{2\gamma_1}{\gamma_1+2}}$&
$\frac{2\sqrt{\pi}C_{\mu 1}^{\frac{2+\gamma_1+2\gamma_2}{2\gamma_1}}{C_2}}
{\left(2+\gamma_1\right){\left(2C_1\right)}^{\frac{1+\gamma_2}{\gamma_1}}}
\frac{\Gamma\left(\frac{1+\gamma_1+\gamma_2}{\gamma_1}\right)}
{\Gamma\left(\frac{2+3\gamma_1+2\gamma_2}{2\gamma_1}\right)}$\\
\hline
\end{tabular}
\caption{Coefficients $C_{\mu 1}$ and $C_{\mu 2}$ of the expansion of the
chemical potential $\mu$ of a trapped system (\ref{mu}) obtained from evaluation of
the normalization integral (\ref{LDANdim}). In a 3D system the integral is to be
taken inside a sphere and can be evaluated explicitly if condition
$(1+\gamma_2)/\gamma_1+1>0$ is satisfied (see, for example, \cite{Gradstein80}). In
a two-dimensional system, $D=2$, the integral is to be calculated inside a circle of
unit radius. There is a substitution
$\int_{0}^{1}f(1-\tilde{r}^2)2\pi\tilde{r}\;d\tilde{r}=$ $\int_{0}^{1}f(y)\pi\;dy$
which together with a particular form of the chemical potential (\ref{muhom}) leads
to an elementary integral. In a one-dimensional system, $D=1$, the spatial
integration over the coordinate $z$ goes from $-R$ to $R$. In contrast to the
two-dimensional case no simplification is found and the final expression for the
chemical potential $\mu$ (\ref{mu}) will contain the gamma functions as in fully
three-dimensional case. The integration was carried out under conditions
$\gamma_1+\gamma_2+1>0$.}
\label{tableCoef}
\end{table}

The energy of the trapped system can be obtained by integrating the chemical
potential $E=\int_{0}^{N}\mu (N')dN'$. From Eqs. (\ref{mu}) and (\ref{Lambda}) we
derive an explicit expression
\begin{eqnarray}
\tilde{E}\equiv\frac{E}{\frac{1}{2}N^{1/D}\hbar \omega_{ho}}
=N\frac{(D\gamma_1+2)C_{\mu 1}}{\left( D+2\right)\gamma_1+2}\Lambda_D^{
\frac{D\gamma_1-2}{D\gamma_1+2}}\left(1+\frac{(D{{\gamma }_{1}}
+2\gamma_1+2)C_{\mu 2}}{\left( D+2\right){{\gamma }_{1}}+2\gamma_2+2
}\Lambda_D^{\frac{2D\gamma_2}{D{{\gamma }_{1}}+2}}+...\right)
\label{E}
\end{eqnarray}

The density profile of a trapped gas can be accessed experimentally by using an
absorption imaging technique, when the shadow of the cloud of the cloud is imaged on
a CCD camera. The size of the cloud can be extracted from such images. In our
approach the size of the cloud along axis $\alpha$ ($\alpha=x,y,z$) is given by
$R_\alpha = R\omega_{ho}/\omega_\alpha$, where the expression for the Thomas-Fermi
radius $R$ is 
\begin{equation}
R = N^{\frac{1}{2D}}C_{\mu1}^{1/2}{\Lambda}_D^{\frac{1}{2}\frac{D\gamma_1-2}{
D\gamma_1+2}}\left(1+\frac{C_{\mu 2}}{2}
\Lambda_D^{\frac{2D\gamma_2}{D\gamma_1+2}}+...\right) a_{ho}
\label{R}
\end{equation}

The applicability of the local density approximation is that the energy of the gas
(\ref{E}) is large compared to the interlevel spacing $E\gg\hbar\omega_{ho}$, so
that the discretization of the levels in the trap can be ignored. The same condition
it terms of length means that the size of the gas (\ref{R}) is large compared to the
oscillator length $R\gg a_{ho}$. These conditions are usually satisfied for a large
number of atoms. In a one-dimensional system the LDA works well even for a small
number of atoms \cite{Astrakharchik02b}, as effects of interactions are enhanced in
low-dimensional systems compared to three-dimensional systems.

The density profile $n(r)$ has a typical shape for the local density approximation:
\begin{equation}
n(r)a_{ho}^{D}=\left[\frac{\tilde{\mu}\Lambda_D}{2C_1}
\left(1-\frac{r^2}{R^2}\right)\right]^{\frac{1}{\gamma_1}}-\frac{C_2}{\gamma_1}
\left[\frac{\tilde{\mu}\Lambda_D}{2C_1}\left(1-\frac{r^2}{R^2}\right)
\right]^{\frac{1+\gamma_2}{\gamma_1}}+...
\label{n}
\end{equation}
where the chemical potential is given by (\ref{mu}) and the size of the cloud is as
in (\ref{R}). For example, for an expansion in the mean-field regime $\gamma_1=1$,
the leading term has the shape of an inverted parabola. In the case of a 1D weakly
interacting Fermi gas or Tonks-Girardeau gas, $\gamma_1=2$, and it takes shape of a
semicircle.

Another important spatial quantity is the mean square displacement $\langle
r^2\rangle$. It is directly related to the potential energy of the harmonic
confinement. The integration of $r^2$ over the density profile (\ref{n}) can be performed
and the result of such an integration is conveniently written in units of the size
of the cloud:
\begin{eqnarray}
\frac{\langle r^2\rangle}{R^2} = \frac{D\gamma_1}{2 + (2 + D)\gamma_1}
\left(1 + \frac{C_2\gamma_2\Gamma(1 + \frac D2 + \frac{1}{\gamma_1})
\Gamma(1+\frac{1 +\gamma_2}{\gamma_1})}
{\gamma_1\Gamma(\frac{1}{\gamma_1})\Gamma(2 + \frac D2 + \frac{1 + \gamma_2}{\gamma_1})}
\left(\frac{\Lambda_D\tilde \mu}{2C_1}\right)^{\frac{\gamma_2}{\gamma_1}}\right)
\label{Rho2}
\end{eqnarray}

Another relevant quantity that can be accessed in experiments is the release energy
defined as the difference of the total energy $E$ of the trapped gas and the
potential energy $E_{ho}=\frac{N}{2}m\omega^2_{ho}\langle r^2\rangle$ 
of the confinement: $E_{rel} = E - E_{ho}$. The potential energy of the confinement
can be eliminated by switching off the trap and the energy of the expanding cloud is
given by $E_{rel}$. Combining together Eqs.~(\ref{E}) and (\ref{Rho2}) and after some
algebra we obtain the following expression for the dimensionless release energy
$\tilde E_{rel} = E_{rel}/\frac{1}{2}N^\frac{1}{D}\hbar\omega_{ho}$:
\begin{eqnarray}
\tilde E_{rel}=N
\frac{2C_{\mu 1}\Lambda_D^{\frac{D\gamma_1-2}{D\gamma_1+2}}}{2+\left(2+D\right)\gamma_1}
\left[1+\left(
\frac{1+\frac{1}{\gamma_1}-
\frac{D\left(\gamma_2-1\right)}{2}}{1+\frac{D}{2}+
\frac{1+\gamma_2}{\gamma_1}}C_{\mu2}
-\frac{D{C_2}\gamma_2\Gamma(1+\frac{D}{2}+\frac{1}{\gamma_1})
\Gamma(1+\frac{1+\gamma_2}{\gamma_1})}{2\Gamma(\frac{1}{\gamma_1})
\Gamma(2+\frac{D}{2}+\frac{1+\gamma_2}{\gamma_1})}
{\left(\frac{C_{\mu 1}}{2{C_1}}\right)}^{\frac{\gamma_2}{\gamma_1}}\right)
\Lambda_D^{\frac{2D\gamma_2}{2+D\gamma_1}}\right]
\end{eqnarray}

\section{Frequencies of collective oscillations\label{secFreq}}

A very important aspect of the LDA approach is that it can be used for the
prediction of the frequencies of the collective oscillations. If oscillations are
generated by displacing the center of the trap, the frequency of the resulting
excitation is determined only by the harmonic confinement and is independent of the
interactions between atoms. Thus, this type of excitation can not be used to probe
the equation of state, but helps to measure the trap's frequencies. Instead, if the
collective mode is excited by changing slightly the frequency of the confinement,
the resulting ``breathing'' oscillation depends on the speed of sound in the gas
$c=\sqrt{n/m~\partial\mu/\partial n}$ and probes directly the equation of state.

If the trap is spherical, then the mean square displacement $\langle r^2\rangle$ can
be used to estimate the frequencies of the collective modes. An important
point is that the breathing mode is naturally excited by a closely related operator
$\sum_ir_i^2$. In this case the sum rules approach can be applied leading to a
compact expression for the frequency of the breathing mode
\begin{eqnarray}
\Omega^2_{sph} = -2 {\langle r^2\rangle}\left/
{\frac{\partial\langle r^2\rangle}{\partial \omega^2}}\right.
\label{OmegaSph}
\end{eqnarray}

This formula was obtained in \cite{Menotti02} while analyzing collective frequencies
of a one-dimensional Lieb-Liniger system. The applicability of
formula~(\ref{OmegaSph}) is not restricted to a one-dimensional system and remains
valid in two- and three- dimensional systems.

Calculation of the derivative of the mean square displacement (\ref{Rho2}) with
respect to the square of the trap frequency leads to the following expression for
the frequency $\Omega$ of the breathing mode in a spherical trap:
\begin{eqnarray}
\frac{\Omega^2_{sph}}{\omega_{ho}^2} =
(2 + D{\gamma_1})\left[1 +\frac{D\gamma_2}{2}\left(C_{\mu 2}
+\frac{C_2\gamma_2\Gamma(1 + \frac{D}{2} + \frac{1}{{\gamma_1}})
\Gamma(1+\frac{1 + \gamma_2}{\gamma_1})}
{\gamma_1\Gamma(\frac{1}{{\gamma_1}})\Gamma(2 + \frac{D}{2} +
\frac{1 + \gamma_2}{\gamma_1})}
{\left(\frac{C_{\mu 1}}{2C_1}\right)}^{\frac{\gamma_2}{\gamma_1}}\right)
\Lambda_D^{\frac{2D\gamma_2}{2 + D\gamma_1}}+...\right]
\label{omega_sph}
\end{eqnarray}

The constant term in the expression for the frequencies (\ref{omega_sph}) depends on
the power of the leading contribution to the homogeneous chemical potential
(\ref{muhom}) and on the dimensionality of the system. The dependence on parameter
$\Lambda_D$ enters only in the subleading term which we are able to obtain with this
perturbative calculation.

An alternative method is needed in order to study collective modes in anisotropic
traps where the formula (\ref{OmegaSph}) is not applicable. As proved in
 \cite{Astrakharchik05}, the {\it scaling approach} allows us to calculate the
frequencies of the collective oscillations with extremely high precision in
anisotropic harmonic confinement. The method is based on using a scaling {\it ansatz}
for the time-evolution of the density profile $n(x,y,z,t)\propto
n(a_1(t)x,a_2(t)y,a_3(t)z)$ and solving the hydrodynamic equations under the assumption
that the amplitude of oscillations is small. The oscillation frequencies $\omega$ in
a 3D elongated anisotropic trap with trap frequencies $(\omega_\perp,
\omega_\perp, \omega_z)$ are given by
\begin{eqnarray}
\Omega^2_{3D}=\frac{1}{4}
\left[2\Xi\omega^2_\perp+(2+\Xi)\omega^2_z
\pm\sqrt{(2\Xi\omega^2_\perp+(2+\Xi)\omega^2_z)^2+16(2-3\Xi)\omega^2_\perp\omega^2_z}\right]
\label{omega3D}
\end{eqnarray}

The natural parameter $\Xi$ is related to the average of the compressibility
$mc^2=n\partial \mu_{\hom}/\partial n$ and the potential energy of the harmonic
oscillator:
$\Xi = D\left\langle mc^2\right\rangle/\left\langle E_{ho}\right\rangle$ \cite{Astrakharchik:PhD,Astrakharchik05}.
We evaluate the parameter $\Xi$ explicitly:
\begin{eqnarray}
\Xi = 2(1 + \gamma_1)
+\frac{2{C_2}\gamma_2(\gamma_1 + \gamma_2)
\Gamma(1+\frac{1 + \gamma_2}{\gamma_1})
\Gamma(2 + \frac{D}{2} + \frac{1}{\gamma_1})}
{\Gamma(\frac{1}{\gamma_1}) \Gamma(2 + \frac{D}{2} +\frac{1 + \gamma_2}{\gamma_1})}
{\left(\frac{C_{\mu 1}}{2{C_1}} \right)}^{\frac{\gamma_2}{\gamma_1}}
\Lambda_D^{\frac{2D\gamma_2}{2 + D\gamma_1}}
\label{Xi}
\end{eqnarray}

From Eq.~(\ref{omega3D}) it follows that the solution in a a spherical trap
$\omega_\perp = \omega_z\equiv\omega_{ho}$ can be simplified. There are two
frequencies: $\Omega^2_{3D}/\omega_{ho}^2=2$ (
center of the mass mode) and $\Omega^2_{3D}/\omega^2_{ho}=3/2~\Xi-1$ (``breathing''
mode). Direct calculation shows that the sum rules prediction for the spherical
trap, Eq.~(\ref{omega_sph}) with $D=3$, coincides with the predictions of
Eq.~(\ref{omega3D}). In an elongated trap, $\omega_z\ne\omega_\perp$, it is possible
to excite axial $\Omega_z$ and radial $\Omega_\perp$ modes separately. In a
``cigar''-shaped trap $\omega_z\ll\omega_\perp$ the frequencies of these modes are
$\Omega_z^2/\omega_z^2=3-2/\Xi$ and $\Omega_\perp^2/\omega_\perp^2=\Xi$
correspondingly. Another possible configuration is a ``pancake'' trap, where
$\omega_z\gg\omega_\perp$. In this situation the frequencies are
$\Omega_z^2/\omega_z^2=1+\Xi/2$ and $\Omega_\perp^2/\omega_\perp^2=6-16/(2+\Xi)$.

We can generalize this scaling method to systems with low dimensionality. We
find that in a two-dimensional system the frequencies of small oscillations are given
by
\begin{eqnarray}
\Omega^2_{2D}=\frac{1}{4}\left[(2+\Xi)(\omega^2_\perp+\omega^2_z)
\pm\sqrt{(2+\Xi)^2(\omega^2_\perp+\omega^2_z)^2-32\Xi\omega^2\omega^2_z}\right]
\label{omega2D}
\end{eqnarray}

In a spherical trap the frequencies are equal to: $\Omega^2_{2D}/\omega_{ho}^2=2$ (
center of mass mode) and $\Omega^2_{2D}/\omega^2_{ho}=\Xi$ (``breathing'' mode).
Again, the frequency of one type of excitation is defined only by the the trap
frequency, while the frequency of the other mode can be used to study the equation of
state of a homogeneous system. The result obtained for a spherical trap is the same
as from Eq.~(\ref{omega_sph}) and $D=2$. In an elongated trap
($\omega_z\ll\omega_\perp$) one obtains the following expressions for the frequencies of
collective excitations in terms of the parameter $\Xi$: $\Omega^2_\perp/\omega^2_\perp =
1+\Xi/2$, $\Omega^2_z/\omega^2_z = 4\Xi/(2+\Xi)$.

There is no way of having an anisotropic trap in a one-dimensional system, so the
result of the scaling approach
\begin{eqnarray}
\Omega^2_{1D}/\omega^2 = 1+\Xi/2
\label{omega1D}
\end{eqnarray}
coincides with the prediction of a spherical trap Eq.~(\ref{omega_sph}) and $D=1$.

It worth noticing that for a given equation of state (\ref{muhom}) the dependence on
the number of particles $N$, $a_{ho}$, $a$ (i.e. the interaction strength) enters
only in the second term. This means that by using only the leading term in
(\ref{muhom}) one obtains no dependence on $\Lambda_{3D}$ in (\ref{Xi}) and it is
necessary to consider the next term to describe changes in the frequency. This
justifies the necessity of our perturbative approach (\ref{muhom}).

\section{Discussion\label{secDiscussion}}

\begin{figure}[tbp]
\begin{center}
\includegraphics*[width=0.4\columnwidth]{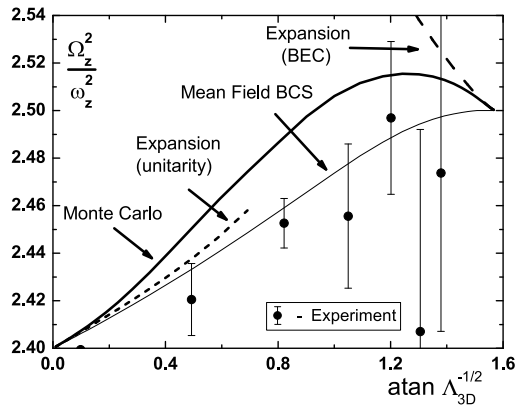}
\end{center}
\caption{Square of the axial mode frequency in ``cigar''-shaped trap for the
Monte Carlo (thick solid line)\cite{Astrakharchik04e} and the mean-field BCS (thin
line) \cite{Leggett80,Nozieres85,Engelbrecht97} equation of states as a function of
$\Lambda_{3D}$. Experimental results (symbols) are taken from \cite{Bartenstein04}.
The dashed lines show the expansions at unitarity (short-dashed line) and in the BEC
regime (long-dashed line) taken from Table~\ref{tableFrequencies3D}.}
\label{Fig1}
\end{figure}

As the frequencies of the collective oscillations can be measured in experiments
with high precision, formulae (\ref{omega_sph}-\ref{Xi}) provide a very useful tool
for the investigation of the equation of state. As an example, the description of
the two-component Fermi gas close to a Feshbach resonance ({\it unitary} regime) is
a very non-trivial problem as there is no obvious small parameter that can be used
to construct an analytically solvable expansion theory (see Fig.~\ref{Fig1}). This
regime has been studied in recent experiments \cite{Kinast04b,Kinast04,Bartenstein04}.
The $s$-wave scattering length in the unitarity regime becomes larger than any other
length-scale in the system $|a_s|\to\infty$ ($\Lambda_{3D}\to\infty$) and,
essentially, it drops out of the problem. The only relevant scale is the density and
as a result the gas has the same scaling behavior as an ideal Fermi gas. This allows
one to fix immediately the leading power $\gamma_1=2/3$, although the next term in
the expansion is not analytically known. It is commonly proposed to be linear in the
parameter $x=-1\;/\,k_Fa_s$ \cite{Combescot04,Bulgac04,Heiselberg04,Kim05}. Thus the
expansion of the chemical potential can be written as $\mu (x)=(\xi + S
x)\hbar^2k_{F}^2/2m$, where the dependence on the density is accounted for by the
Fermi momentum $k_{F}=(3\pi^2n)^{1/3}$. We find that the predictions of the sum-rule
approach (see Table~\ref{tableFrequencies3D}) coincide in the unitary regime with
the results of \cite{Combescot04}, where the
authors start directly from the hydrodynamic equations and solve them for this
particular choice of the equation of state. The second row of the
Table~\ref{tableFrequencies3D} shows the predictions in the
Bardeen-Cooper-Schrieffer (BCS) limit $a_s\to -0$ ($\Lambda_{3D}\to 0$), where fermions
experience a weak attraction and the equation of state can be found perturbatively
starting from a weakly interacting Fermi gas. The leading term corresponds to an
ideal Fermi gas and scales as $n^{2/3}$ with the density. This leads to the same
leading term in the frequencies of the collective oscillations as at unitarity,
although the next term \cite{Huang57,Lee57} is different and arises from the interparticle
interaction of a normal Fermi gas.

The situation in the Bose-Einstein condensation (BEC) regime $a_s\to +0$
($\Lambda_{3D}\to 0$) is extremely interesting. In this limit a two-body bound state
(molecules) of atoms of different spin is formed. The main contribution to the
energy comes from the negative binding energy of molecules. As the binding energy
depends only on the interaction potential and is independent of the density limit,
it does not contribute to the collective excitations and this term is omitted in
Table~\ref{tableFrequencies3D}. The first leading term depending on the density
corresponds to a dilute non-ideal Bose gas of molecules, which play the role of
composite bosons with mass $2m$ and density $n/2$. The $s$-wave scattering length
$a_m$ of such composite bosons was found directly by solving the four-body
scattering problem \cite{Petrov04} and was estimated to be $a_m=0.6\;a_s$. It should
be noted that this result differs from the commonly used mean-field BCS theory which
gives $a_m=2\;a_s$ (see, for example, \cite{Leggett80,Nozieres85,Engelbrecht97}).
The nature of the the next term in the equation of state is still an open question
and only a precise experimental measurement can give a definitive answer. This term
might have a bosonic nature and correspond to the beyond mean-field correction
 \cite{Lee57b} to the equation of state of composite bosons as proposed in
 \cite{Pitaevskii98} (see, also, \cite{Manini05}). Alternatively, the next correction
might be of a fermionic nature as argued in \cite{Combescot04b}. An {\it ab initio}
quantum Monte Carlo study of the many-body problem made it possible to obtain the equation of
state numerically \cite{Astrakharchik04e}. Firstly, it was shown that the leading
term has a mean-field nature and agrees with a molecule-molecule scattering length
equal to $a_m=0.6\;a_s$ (very recently this result was also recovered from a
diagrammatic approach \cite{Brodsky05}). Secondly, the Monte Carlo study suggests the
presence of the bosonic beyond mean-field correction. As inferred from
Table~\ref{tableFrequencies3D} and shown in Fig.~\ref{Fig1} this would mean that the
frequencies of the collective oscillation should increase above the mean-field value
by making the interaction between molecules stronger. To date there has been no
experimental corroboration of this result. In the available experimental work \cite{Bartenstein04}
the frequency decreases, but the deep BEC regime has not yet been achieved. A more
precise experimental study should be performed, in which case Eq.~(\ref{Xi}) can be used
to fit the frequencies of the collective oscillations and thus provide insight into the
nature of the equation of state.

We provide a relationship between the characteristic LDA parameter $\Lambda_{3D}$
and the value of the Fermi momentum in the center of the trap $k_{F,max}$, which is
a natural quantity for a fermionic system. The leading term (which is sufficient for the
applicability of Eq.~(\ref{Xi})) is given by
\begin{eqnarray}
\Lambda_{3D}=(2\pi C_1)^{1/2}{\left(\frac{\Gamma\left(1+\frac{1}{\gamma_1}\right)}
{\Gamma\left(\frac{5}{2}+\frac{1}{\gamma_1}\right)}\right)}^{1/3}
\left(\frac{k_{F,max}^3}{3\pi^2}\right)^{\frac{1}{3}+\frac{\gamma_1}{2}}
\end{eqnarray}

\begin{table}[ht!]
\centering
\begin{tabular}{|c|l|c|l|r|l|l|l|}
\hline
Limit & $C_1$ & $\gamma_1$ & $C_2$ & $\gamma_2$ & $\Omega_\perp^2/\omega_\perp^2$ & $\Omega_z^2/\omega_z^2$ &$\Omega_{sph}^2/\omega_z^2$\\ \hline
BCS & $\displaystyle\frac{(3\pi^2)^{2/3}}{2}$ & $\displaystyle\frac{2}{3}$
& $\displaystyle\frac{4}{3^{2/3}{\pi}^{1/3}}$ & $\displaystyle\frac{1}{3}$
& $\displaystyle\frac{10}{3} + \frac{8192\sqrt{2}\Lambda_{3D}^{1/2}}{945\,3^{5/6}\pi^2}$
& $\displaystyle\frac{12}{5} + \frac{4096\sqrt{2}\Lambda_{3D}^{1/2}}{2625\,3^{5/6}\pi^2}$
& $\displaystyle 4 + \frac{4096\sqrt{2}\Lambda_{3D}^{1/2}}{315\,3^{5/6}\pi^2}$
\\ \hline
unitary & $\displaystyle\frac{(3\pi^2)^{2/3}\xi}{2}$
& $\displaystyle\frac{2}{3}$
& $\displaystyle-\frac{S}{(3\pi^2)^{1/3}\xi}$ & $\displaystyle-\frac{1}{3}$
& $\displaystyle\frac{10}{3}+\frac{256\sqrt{2}S}{3^{1/6}315\pi\xi^{3/4}\Lambda_{3D}^{1/2}}$
& $\displaystyle\frac{12}{5}+\frac{128\sqrt{2}S}{3^{1/6}875\pi\xi^{3/4}\Lambda^{1/2}_{3D}}$
& $\displaystyle 4 + \frac{128\sqrt{2}S}{3^{1/6}105\pi\xi^{3/4}\Lambda_{3D}^{1/2}}$
\\ \hline
BEC & $\displaystyle\frac{1}{2}\pi a_m$
& $\displaystyle 1$
& $\displaystyle\frac{16\sqrt{2}a_m^{3/2}}{3\sqrt{\pi}}$
& $\displaystyle\frac{1}{2}$
& $\displaystyle 4 + \frac{{15}^{1/5}105a_m^{6/5}\Lambda_{3D}^{3/5}}{2^{1/10}64}$
& $\displaystyle \frac{5}{2} + \frac{{15}^{1/5}105a_m^{6/5}\Lambda_{3D}^{3/5}}{2^{1/10}512}$
& $\displaystyle 5 + \frac{{15}^{1/5}315a_m^{6/5}\Lambda_{3D}^{3/5}}{2^{1/10}128}$
\\ \hline
Bose gas & $\displaystyle 4\pi$
& $\displaystyle 1$
& $\displaystyle\frac{32}{3\sqrt{\pi}}$
& $\displaystyle\frac{1}{2}$
& $\displaystyle 4 + \frac{{15}^{1/5}105\Lambda_{3D}^{3/5}}{64\sqrt{2}}$
& $\displaystyle \frac{5}{2} + \frac{{15}^{1/5}105\Lambda_{3D}^{3/5}}{512\sqrt{2}}$
& $\displaystyle 5 + \frac{{15}^{1/5}315\Lambda_{3D}^{3/5}}{128\sqrt{2}}$
\\ \hline
\end{tabular}
\caption{Summary for three-dimensional systems. The first column labels the physical system under
considerations: limits for the two-component Fermi gas in the BEC-BCS crossover and
weakly-interacting system of bosons. Columns 2-5: coefficients of the expansion of
the equation of state defined as in Eq.~(\ref{muhom}). Columns 6-7: frequencies of
radial and axial oscillations in a ``cigar''-shaped trap (note that in a 3D system
the parameter (\ref{Xi}) is $\Xi=\Omega_\perp^2/\omega^2_\perp$). Column 8:
frequency of the breathing mode in a spherical trap. The notation $a_m$ in the BEC
regime denotes the molecule-molecule scattering length in units of the atomic scattering
length. In the same regime we omit the contribution to the chemical potential
from the binding energy, as it is independent of the density and, thus, is not
important for the LDA.}
\label{tableFrequencies3D}
\end{table}

Low-dimensional geometries are advantageous for the investigation of the energetic
properties of a gas as the reduced phase space enhances the role of interactions. As
a result, the frequencies of collective oscillations vary over a larger range, thus
facilitating possible experimental measurements. We consider a quasi-two-dimensional
system of bosons in the regime of small densities. We assume that the gas is weakly
bound in two-directions $(x,y)$ and is subjected to a tight confinement in $z$
direction. The density profile in a uniform system can be approximated as
$n(x,y,z)=n_{2D}\exp(-z^2)/\sqrt{\pi a_{osc}^2}$, where $n_{2D}$ is the
two-dimensional density. Locally the equation of a rarefied three-dimensional bose gas
can be applied (see Table~\ref{tableFrequencies3D}) and the energy of the gas is
estimated by integrating out the variable corresponding to the tight confinement
$E_{2D}(n_{2D}) = \int E_{3D}(n(x,y,z)) n(x,y,z) dz/\int n(x,y,z) dz$. The chemical
potential, obtained by differentiating the energy, is reported in
Table~\ref{tableFrequencies2D}. We calculate frequencies of collective oscillations
by exploiting Eqs.~(\ref{Xi}-\ref{omega2D}). The results for a spherical and very
elongated traps are presented in Table~\ref{tableFrequencies2D}. We note that the
quasi two-dimensional equation of state is valid when the scattering length $a_s$ is
small compared to the oscillator length of the tight confinement, so that the
scattering process is still three-dimensional. In a ``pure'' two-dimensional system
the equation of state contains logarithmic terms \cite{Schick71} and our approach
does not apply.

\begin{table}[ht!]
\centering
\begin{tabular}{|c|c|c|c|c|c|c|c|}
\hline
system&$C_1$ & $\gamma_1$ & $C_2$ & $\gamma_2$ & $\Omega_\perp^2/\omega_\perp^2$ & $\Omega_z^2/\omega_z^2$ &$\Omega_{sph}^2/\omega_z^2$\\
\hline
Q2D Bose gas
& $\displaystyle 2\sqrt{2\pi}$
& $\displaystyle 1$
& $\displaystyle \frac{64}{3\sqrt{5}\pi^{3/4}}$
& $\displaystyle 1/2$
& $\displaystyle 3 + \frac{2^{5/8}192\Lambda_{2D}^{1/2}}{35\sqrt{5}\pi^{9/8}}$
& $\displaystyle \frac{8}{3} + \frac{2^{5/8}256\Lambda_{2D}^{1/2}}{105\sqrt{5}\pi^{9/8}}$
& $\displaystyle 4 + \frac{2^{5/8}384\Lambda_{2D}^{1/2}}{35\sqrt{5}\pi^{9/8}}$
\\
\hline
\end{tabular}
\caption{Equation of state of a quasi-two-dimensional bose gas in the weakly
interacting regime. The expansion parameter is $n_{2D}a^3/a_{osc}$. It is assumed
that $a\ll a_{osc}$. The meaning of the columns is the same as in
Table~\ref{tableFrequencies3D}.}
\label{tableFrequencies2D}
\end{table}

In Table~\ref{tableFrequencies1D} we summarize the examples of bosonic as well as
fermionic systems in (quasi-) one-dimensional geometry. The Lieb-Liniger model
 \cite{Lieb63} describes a gas of repulsive bosons interacting with a contact
$\delta$-potential. The equation of state of such a gas can be expanded in the
weakly interacting limit $\Lambda_{1D}=Na_{1D}^2/a_{ho}^2\to\infty$ (one notes that
in a waveguide small values of the three-dimensional $s$-wave scattering length
$a_{3D}$ correspond to small values of the one-dimensional $s$-wave scattering length
$a_{1D}$ as in this limit $a_{1D}=-a_\perp^2/a_z$). The leading term in the chemical
potential $\mu$ is linear in the density and corresponds to the mean-field
Gross-Pitaevskii theory. Zero point oscillations of quasi-particles contribute to
the next term and in a three-dimensional system corresponds to the Lee-Yang correction
\cite{Lee57}. We applied the Bogoliubov theory to a one-dimensional system and found
that the corresponding term coincides with the second term of the expansion of
Lie-Liniger equation of state. In the opposite limit of strong correlations
$\Lambda_{1D}\to 0$ (Tonks-Girardeau regime) the leading term in $\mu$ equals the
Fermi energy of ideal spinless fermions due to a very peculiar property of a
Tonks-Girardeau gas experiencing ``fermionization'', as the wavefunction of
impenetrable point-like bosons can be mapped directly onto a wavefunction of a
fermionic system \cite{Girardeau60}. Another one-dimensional system, the
Calogero-Sutherland gas \cite{Calogero69,Sutherland71} has a functionally similar
equation of state (see Table~\ref{tableFrequencies1D}, parameter $\lambda$ is
related to the strength of the $1/|z|^3$ interaction). In order to understand the
physical meaning of the subleading correction we will consider for a moment a system
of hard-rods, {\it i.e.} impenetrable bosons of a size $|a_{1D}|$. Its equation of
state can be obtained from the energy of a Tonks-Girardeau gas by reducing the
density by the excluded volume $n\to n-N|a_{1D}|$. The leading term is again given
by the energy of an ideal Fermi gas of spinless particles and the next correction is
given in Table~\ref{tableFrequencies1D}. It was shown in
 \cite{Astrakharchik04d} that a lowest gas-like state of attractive bosons
($a_{1D}>0$) in the limit of small $a_{1D}$ (super-Tonks gas) has the same leading
term in the expansion of the equation of state as a system of hard-rods. The reason
is that the node of a two-body low-energy scattering solution lies at $a_{1D}$ and
for larger distances is essentially the same as in a system of hard-rods. Instead,
for a repulsive gas with a contact interaction in the vicinity of the
Tonks-Girardeau limit, the analytic continuation of the scattering solution has a
node at $-|a_{1D}|$. One finds that the subleading term in the equation of state is
equal to the ``excluded volume'' correction, but differs in sign (see
Table~\ref{tableFrequencies1D}). Of course, the discussed terms in the chemical
potential of the Lieb-Liniger gas coincides with the ones found by solving
iteratively the Bethe {\it ansatz} integral equations. Still we find it important to
understand the physical origin of the terms of the expansion.

\begin{table}[ht!]
\centering
\begin{tabular}{|l|l|l|l|r|l|}
\hline
Limit & $C_1$ & $\gamma_1$ & $C_2$ & $\gamma_2$ & $\Omega_z^2/\omega_z^2$ \\
\hline
Lieb-Liniger: weak interaction & $2$ & $1$ & $-\sqrt{2}/\pi$ & -1/2 & $
\displaystyle3+\frac{3^{2/3}5}{32\sqrt{2}}/{\Lambda_{1D}^{1/3}} $ \\ \hline
Lieb-Liniger: strong interaction & $\pi^2/2$ & 2 & -8/3 & 1 & $\displaystyle
4-\frac{128\sqrt{2}}{15\pi^2}{\Lambda^{1/2}_{1D}}$ \\ \hline
Attractive Fermi gas: strong interaction & $\pi^2/32$ & 2 & 2/3 & 1 & $
\displaystyle4+\frac{64\sqrt{2}}{15\pi^2}{\Lambda^{1/2}_{1D}}$ \\ \hline
Attractive Fermi gas: weak interaction & $\pi^2/8$ & 2 & $-8/\pi^2$ & -1 & $
\displaystyle 4+\frac{32}{3\pi^2}/{\Lambda^{1/2}_{1D}}$ \\ \hline
Repulsive Fermi gas: strong interaction & $\pi^2/2$ & 2 & $-8\ln (2)/3$ & 1
& $\displaystyle4-\frac{128\sqrt{2}\ln 2}{15\pi^2}{\Lambda^{1/2}_{1D}}$ \\ \hline
Repulsive Fermi gas: weak interaction & $\pi^2/8$ & 2 & $8/\pi^2$ & -1 & $
\displaystyle4-\frac{32}{3\pi^2}/{\Lambda^{1/2}_{1D}}$ \\ \hline
Gas of Hard-Rods & $\pi^2/2$ & 2 & 8/3 & 1 & $\displaystyle 4+\frac{128\sqrt{
2}}{15\pi^2}{\Lambda^{1/2}_{1D}}$ \\ \hline
Calogero-Sutherland & $\lambda^2\pi^2/2$ & 2 & 8/3 & 1 & $\displaystyle 4+\frac{128\sqrt{
2}}{15\pi^2\sqrt{\lambda}}{\Lambda^{1/2}_{1D}}$ \\ \hline
\end{tabular}
\caption{Summary of some of the one-dimensional models where the expansion of
the equation of state is known. The first column names the considered systems. The
coefficients of the expansion are given in columns 2-5. The last column gives the
predictions (\ref{omega1D}) for the oscillation frequencies. The parameter
$\Lambda_{1D}$ is defined by (\ref{Lambda}). Note that the presence of a term in the
chemical potential independent of the density ({\it e.g.} binding energy of a
molecule) does not modify the frequencies of oscillations and is ignored.}
\label{tableFrequencies1D}
\end{table}

Considering a system of one-dimensional fermions and limiting ourselves to the
$s$-wave scattering we immediately find that atoms of the same spin do not
interact due to the Pauli exclusion principle. Instead, fermions of different spin are
allowed to interact and the interaction can be either attractive or repulsive.
The equation of state can be found by solving the integral Bethe {\it ansatz}
equations as discussed for attractive \cite{Gaudin67,Krivnov75} and repulsive
 \cite{Yang67} interactions. The numerical solution for the LDA was carried out in
 \cite{Astrakharchik04b} and is compared here to the series expansions obtained from
Eq.~(\ref{omega1D}). Again, there is interesting physics behind the expansion
(\ref{mu}). In the weakly interacting limit (both attraction and repulsion) the
leading term equals the energy of an ideal two-component Fermi gas. The next
correction is linear in density and describe the mean-field interactions between
particles of different spin. In the limit of strong repulsion the main
contribution to $\mu$ coincides with the chemical potential of a {\it one-component}
ideal 1D Fermi gas with $N$ atoms, as the strong repulsion between atoms of
different spin plays the role of an effective Pauli principle \cite{Recati03}.
An attractive interaction in one-dimension immediately leads to the formation of a bound
state (molecule). As the binding energy is independent of the density and does not
contribute to the collective oscillations, this term is omitted in
Table~\ref{tableFrequencies3D}. In the low-density limit the internal structure of a
molecule of two fermions can be neglected and be considered as a composite boson. Since
in this limit the interaction between such composite bosons is repulsive and strong
(Tonks-Girardeau limit of $N/2$ bosons of mass $2m$), the system can be mapped to
$N/2$ spinless fermions. So the leading behavior is again of a fermionic nature. The
next term was obtained correctly for the first time in
 \cite{Astrakharchik:PhD,Astrakharchik04b} and is reported in
Table~\ref{tableFrequencies1D}.

\begin{figure}[tbp]
\begin{center}
\includegraphics*[width=0.4\columnwidth]{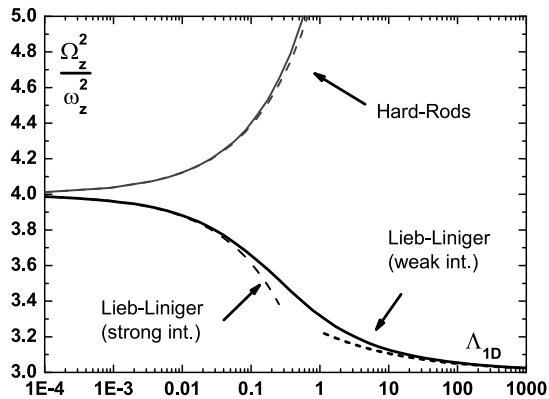}
\end{center}
\caption{Frequency of the breathing mode in a number of one-dimensional
bosonic systems. The frequencies for the Lieb-Liniger model were obtained
numerically in \cite{Menotti02}. The collective excitations of a gas of hard-rods
were discussed in \cite{Astrakharchik04d} in their relation to the excitations of a
gas in the super-Tonks regime. The dashed lines correspond to the expressions given
in Table~\ref{tableFrequencies1D}}
\label{Fig1Dbosons}
\end{figure}

\begin{figure}[tbp]
\begin{center}
\includegraphics*[width=0.4\columnwidth]{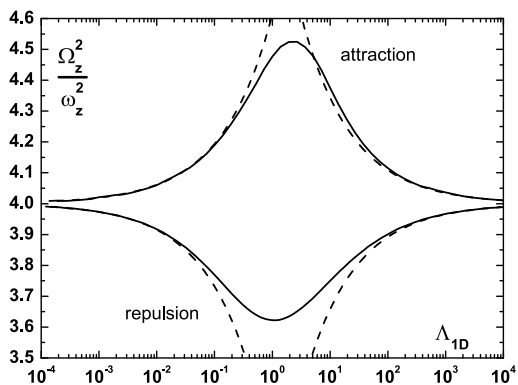}
\end{center}
\caption{Frequencies of the breathing mode of a two-component Fermi gas in a
quasi one-dimensional geometry. The solid line shows results obtained by solving
numerically the LDA with an exact equation of state \cite{Astrakharchik04b}. The
dashed lines show the small- and large- density expansions from
Table~\ref{tableFrequencies1D}. The upper curve corresponds to attraction between
atoms of opposite spin, while the lower curve corresponds to repulsion.}
\label{Fig1Dfermions}
\end{figure}

In Figs.~\ref{Fig1Dbosons},~\ref{Fig1Dfermions} we visualize how the terms of the
EOS expansion show up in the frequencies of collective modes. The
Fig.~\ref{Fig1Dbosons} shows the frequencies for bosonic systems, although the
Fig.~\ref{Fig1Dfermions} presents results for systems of two-component fermions. We
compare the predictions of the expansion, Eq.~(\ref{omega1D}), (dashed lines) with
the numerical solution for the exact equation of state (solid lines). As it can be seen
from the Figures, the region where the perturbative results hold is quite large,
making possible the extraction of the expansion terms of the equation of state from
the frequencies of collective oscillations. The accuracy is better for the
expansions at low density, since if the expansion holds for the center of the trap it
will certainly be valid close to the borders of the trap where the density is
smaller (see, {\it e.g.} Fig.~\ref{Fig1Dfermions}).

\section{Conclusions\label{secConclusions}}

It quite often happens that a perturbation theory for a homogeneous system can be developed
analytically, although no solution for a system in an external confinement is known.
If the number of particles in a trapped system is large enough (and this requirement
often can be easily fulfilled in experiments) the local density approximation can be
applied. We assume that the series expansion of the dependence of the chemical
potential on density is known for a homogeneous system and we find the properties of a
gas confined in an anisotropic trap. We consider three-, two-, and one- dimensional
systems. We obtain the energetic properties such as the chemical potential, total
energy of the gas, potential energy of the confinement. The calculated release
energy can be accessed in expansion experiments. Spatial properties such as the
density profile and size of the cloud are predicted and can be measured by imaging
techniques in current experiments.

The frequencies of the collective oscillations in a harmonic trap are predicted from
a hydrodynamic theory. We apply a scaling approach to obtain frequencies in three-,
two-, and one- dimensional systems in traps with arbitrary aspect ratio
$\omega_\perp/\omega_z$. In spherical, ``cigar''- and ``pancake''- shaped geometries
the expressions are simplified. In a spherical trap an equivalent method, the
sum-rule approach, can be conveniently applied and leads to analogous expressions.
Exploiting the obtained expressions and using our solution to the LDA problem we relate the
frequencies of the breathing modes to the expansion of the equation of state of a
homogeneous systems in an explicit way.

We argue that the obtained results are highly relevant for the experimental study of the
equations of state of trapped gases. The expansion of the equation of state can be
found by a best fit of our formulas to experimentally measured quantities (for
example, collective oscillation frequencies). This would provide an important
insight into the physics of the systems under the study. As an example, there is an
open and intriguing question of what will be the nature of the beyond-mean field
term in the equation of state of the recently observed condensation of molecules. Will
it have a bosonic or fermionic nature? Both proposals can be found in the literature
and only a high precision experiment can provide a definitive result.

We apply the derived formalism to a number of three-dimensional systems
(two-component Fermi gas in the BEC-BCS crossover, dilute Bose gas), quasi
two-dimensional bose gas, quasi one-dimensional systems (Lieb-Liniger gas, gas of
hard-rods, two-component fermionic gas, Calogero-Sutherlanf gas). A comparison with
numerical solutions is provided. Finally we provide an physical interpretation of the
expansion terms in a number of one-dimensional systems.

This work was supported by Ministero dell'Istruzione, dell'Universit\`a e della
Ricerca (MIUR). Useful discussions with S. Giorgini and S. Stringari are gratefully
acknowledged. We thank B.~Jackson for reading the manuscript.

\end{document}